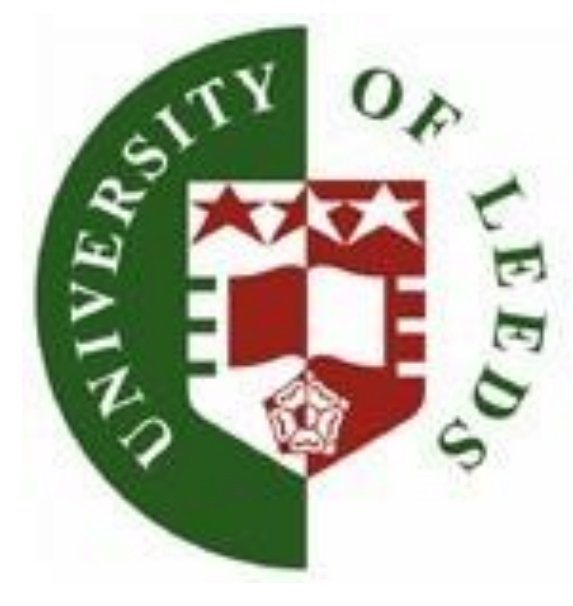

UNIVERSITY OF LEEDS

Faculty of Mathematics and Physical Sciences

Department of Physics and Astronomy

# ACTIVE GALACTIC NUCLEI

Submitted by: Yerbol Khassen, ID No. 200299063

Supervised by: Dr. Stuart Lumsden

Program of study: BSc Physics with Astrophysics

# ACKNOWLEDGEMENTS

My sincere thanks to my personal tutor Dr. Joachim Rose for his huge support and guidance. I also wish to thank Dr. Bradbury for her advice and help during the period of my studies. I would like to mention all the staff and colleagues in the Department for creating friendly atmosphere which positively influenced my academic life. Special thanks to Mr. Everton for answering all my unlimited queries. I also cherish the emotional support from my family members and best friends. Further, I convey my special thanks to Dr Stuart Lumsden who supervised my work and for his invaluable support during this project.

## ABSTRACT

This work represents the final year project for BSc Physics with Astrophysics degree and it mainly focuses on empirical investigation of the photometry of quasars in the Sloan Digital Sky Survey (SDSS) and the UK Infrared Telescope (UKIRT) Infrared Sky Survey (UKIDSS) systems. The studies include 5730 quasars matched from both surveys and examine UV/optical/near-IR properties of the population. The sample covers the redshift and absolute magnitude ranges $0.01 < z < 3$ and $-29.3 < M_i < -13.8$ and 17 per cent of the SDSS quasars have matching success to the UKIDSS data. The combination of SDSS *ugriz* with the JHK near-IR photometry from UKIDSS over large areas of the sky has enormous potential for advancing our understanding of quasar population, keeping in mind that these surveys have not reached their terminations.

# 1. INTRODUCTION

## *1.1. Structure of AGN*

Active Galactic Nuclei (AGN) have been studied intensely in all portions of the electromagnetic spectrum since the mid-20th century. Both ground and satellite-based astronomy have provided wealthy amount of data ranging from radio to gamma rays. Better and ever-improving facilities in tact with advances in theoretical modelling provided clear understanding of the physics behind AGN [15]. However, the effects of AGN to its galaxy formation and evolution still require more detailed investigations [5].

The general consensus today between astronomers is that a small portion of galaxy population contains very powerful non-stellar energy sources called AGN [18]. It is believed that the activity is driven by the accretion of gas to supermassive black holes (SMBH) at the centres of galaxy [20]. Accretion disk is a result of matter in the host galaxy falling towards SMBH due to loss of angular momentum. As the gas falls inwards, the gravitational energy is converted into heat and light. Consequently, we observe objects with wide range of luminosities and variability. Thus, AGN are classified as low-ionisation nuclear emission-line region (LINER), Seyfert galaxies, quasi-stellar objects (QSO), blazing quasars (blazars), BL LAC objects, optically violent variable (OOV) quasars and radio galaxies [18].

Furthermore, AGN are divided into two types characterised by the spectral feature of broad emission lines emanating from broad line region (BLR). Type 1s have Doppler-broadened permitted lines with full width half maximum (FWHM) > 1000 km/s, whereas Type 2s show permitted and forbidden lines of similar widths. Today it is predominantly agreed upon unified model which theorises that the variety in AGN types is a result of different orientations relative to the line of sight [4]. The main concept of unification is that two types are intrinsically the same but BLR is blocked by optically thick torus in Type 2s (Figure 1).

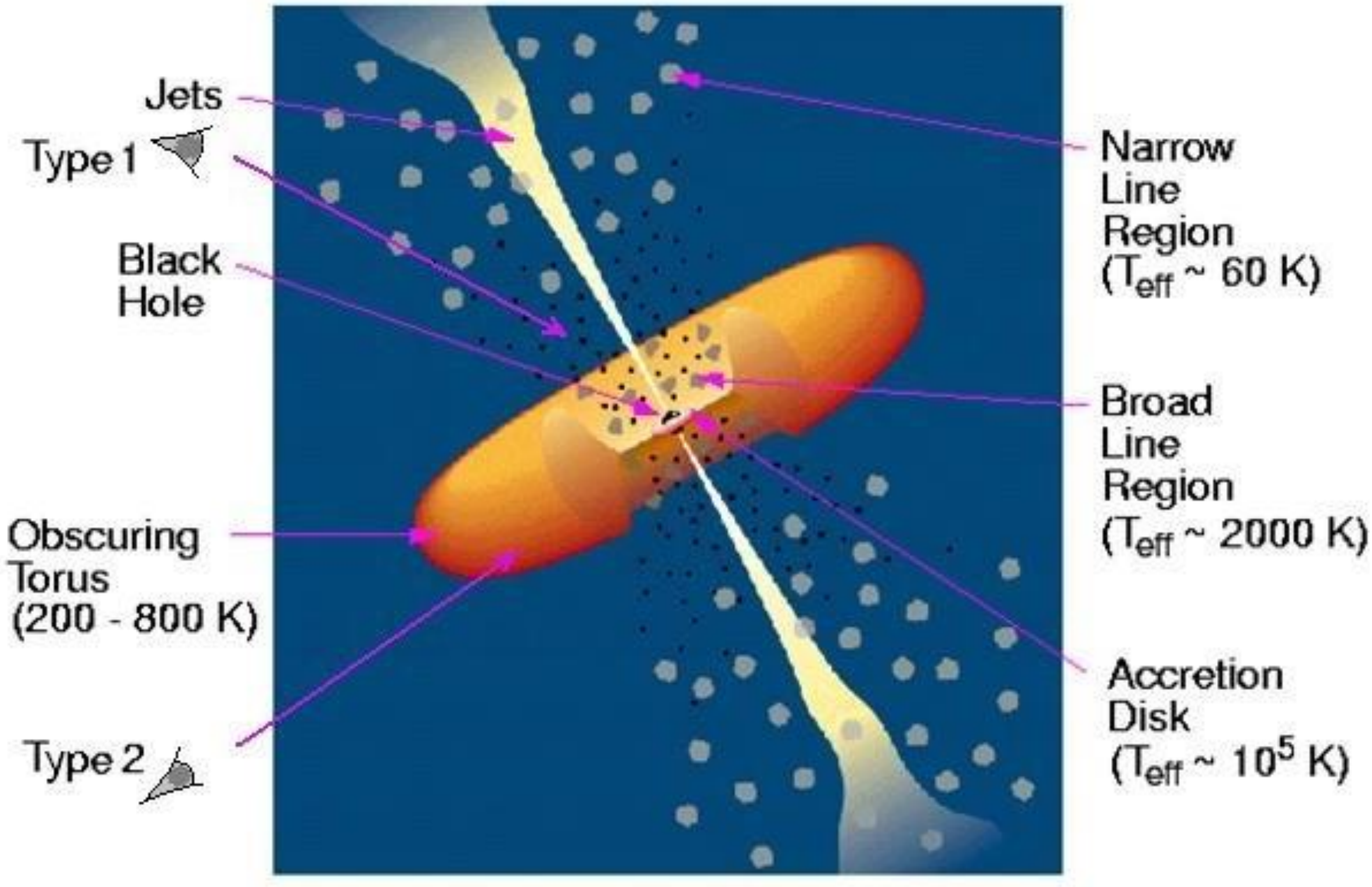

**Figure 1.** Schematic diagram of the current model for AGN (not to scale). Surrounding the central SMBH is a luminous accretion disk. Broad emission lines are produced in clouds orbiting above the disk and perhaps by the disk itself. Narrow lines are generated in clouds much further from the central source. AGN are not spherically symmetric and thus what you see depends on from where you view them. This is the basis of most unification models (Source: [31]).

### *1.2. Importance of quasars*

This work focuses mostly on quasi-stellar objects, or simply quasars, because they are the brightest objects found in the universe to date [25]. Highly luminous quasars may serve as crucial cosmological probes, since they can be detected at large distances. Another critical role of quasars lies in the evolution of the total star formation rate density of the universe [17]. There is increasing evidence that the space density of starbursting galaxies and that of luminous quasars coincide remarkably. According to Peterson (1997), one of the defining characteristics of quasars is their very broad spectral energy distribution, or SED. It is dominated by non-thermal processes such as incoherent synchrotron radiation and can be described in a power law

$$F_\nu = C\nu^{-\alpha}$$

where α is the power law index, $C$ is a constant and $F_\nu$ is the specific flux in $[erg\, s^{-1} cm^{-2} Hz^{-1}]$.

The data used in this work has been extracted from two large multicolour quasar surveys currently running world-wide, which are SDSS [34] and UKIDSS [19]. Spectroscopically identified quasars from SDSS Data Release (DR) 7 [2] have been positionally cross-matched with UKIDSS DR6 [32] (Figure 2).

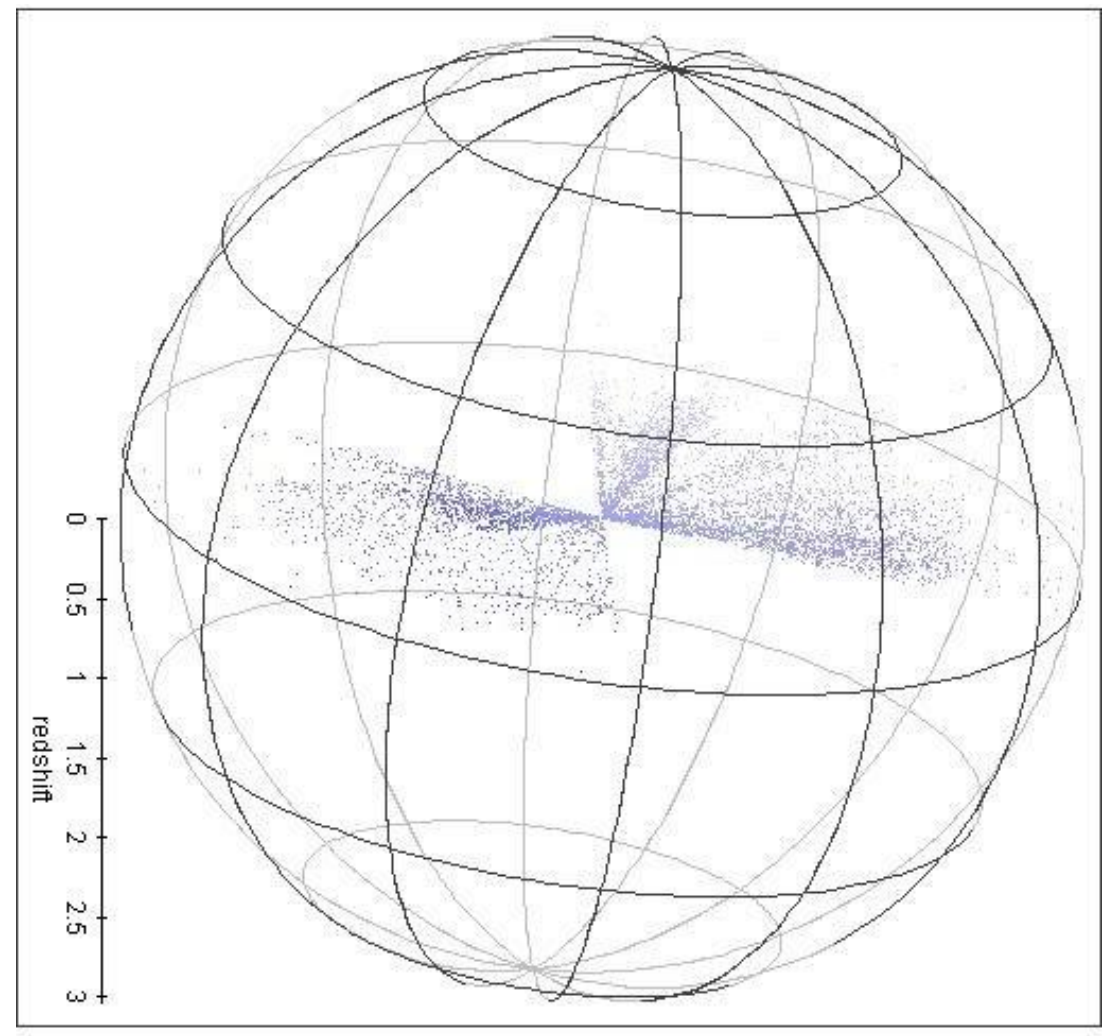

**Figure 2.** 3D map of SDSS DR7 quasars matched with UKIDSS DR5 are plotted in sky coordinates (J2000) with distance from the centre as a redshift. Most quasars were found near the galactic equator (Stripe 82 of SDSS).

Combination of the SDSS optical and UKIDSS near-infrared data allows more comprehensive insight into the photometric and spectroscopic properties of quasars relative to the quasar population as a whole [21]. All absolute magnitude calculations were based on the Friedmann-Robertson-Walker cosmology with
$\Omega_M = 0.27,\ \Omega_\Lambda = 0.73$ and $H_0 = 71\ km\,s^{-1}\ Mpc^{-1}$.

## 2. DATA SOURCES

### *2.1. SDSS*

SDSS will map a quarter of the entire sky and perform a redshift survey of million galaxies and 100'000 quasars [34]. The survey uses a dedicated wide-field 2.5m telescope [13] at the Apache Point Observatory (APO) which is equipped with a matrix type CCD camera [12] and a couple of fibre-fed spectrographs. The detection system acquires signal through five simultaneous filters (*u, g, r, i,* and *z* [10]) with average wavelengths of 3551, 4686, 6165, 7481 and 8931Å and magnitude limits of 20.0, 22.2, 22.2, 21.3 and 20.5 respectively (signal-to-noise > 4 per pixel at g = 20.2). SDSS magnitudes are based on the AB system [24] and its calibration of zero points has accuracy of 1-2% [1]. The sensitivity curves of the five bands are shown on Figure 5 with 5σ accuracy cut.

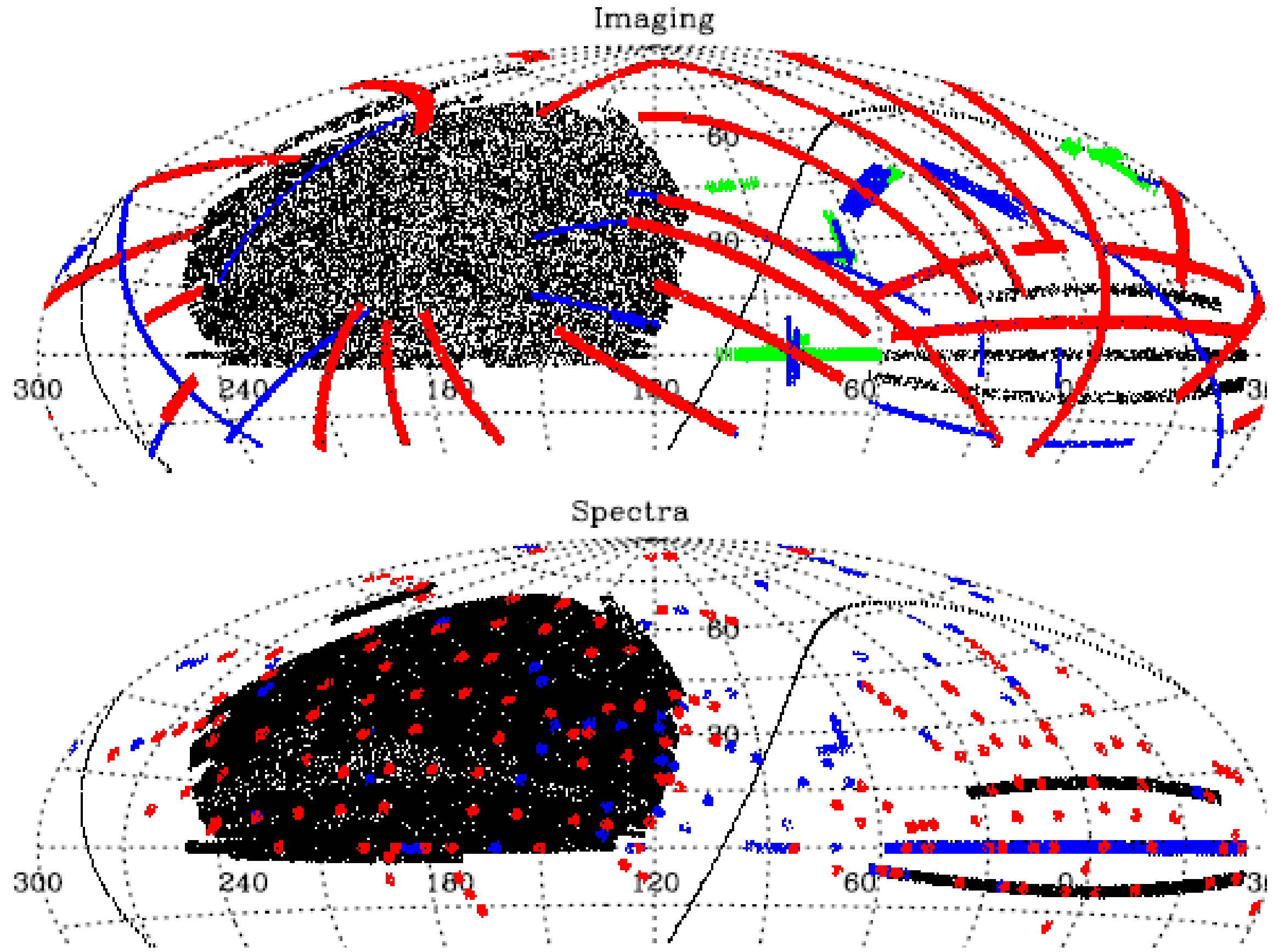

**Figure 3.** Distribution on the sky of the data included in DR7 (upper panel: imaging; lower panel: spectra) show in an Aitoff equal-area projection in J2000 Equatorial Coordinates. The Galactic plane is the sinuous line that goes through each panel. The centre of each panel is at $\alpha = 120° \equiv 8^h$, and the plots cut off at $\delta = -25°$, below which the SDSS did not extend. The Legacy imaging surveys the contiguous area of the Northern Galactic Cap (centred roughly at α=200°, δ=30°), as well as three stripes (each of width 2°.5) in the Southern Galactic Cap. In addition, several stripes (indicated in blue in the imaging data) are auxiliary imaging data, while the SEGUE imaging scans are indicated in red. The green scans are additional runs as described in Finbeiner et al. (2004). In the spectroscopy panel, the lighter regions indicate that area in the Northern Galactic Cap which is new to DR7; note that the Northern Galactic Cap is now contiguous. Red points indicate SEGUE plates and blue points indicate non-Legacy plates (mostly as described in the DR4 paper) (Source: [2]).

The redshift constraints are set to between 0 and 3 for the purpose of this project. Quasar population at this range is thought to be sufficient to test and compare current models and statistical distributions, whereas higher redshift QSO are tend to have significantly shifted profiles (spectrum) and do not essentially affect the population because there are few currently discovered. Abrusco [3] found that one possible problem for quasars of this range is likely to be A stars which are the objects that are strongly affected by the Balmer decrement and these are recognised as the prime contaminants in multicolour optical searches for low-redshifted quasars.

SDSS quasar target selection works through these steps (more detailed explanation of the selection code can be found in Richards et al. (2002)):

1) Objects with spurious and/or problematic fluxes in the imaging data are rejected
2) Point-source matches to FIRST radio sources are preferentially targeted without reference to their colours
3) The sources remaining after the first step are compared to the distribution of normal stars and galaxies in two distinct 3D colour spaces, one normally for low-redshift and another for high-redshift quasar candidates.

In other words, the colour selection process provides no definite boundary between quasars and less luminous Seyfert galaxies, so that even extended objects could be classified as quasars. Therefore, in this paper the word "quasar", in fact, implies "AGN". However it is possible to select the true quasars by using boundaries similar to the ones used by Schneider et al. (2005), where they had absolute luminosities brighter than $M_i = -22.0$ and at least one emission line with FWHM larger than 1000 km/s or highly reliable redshifts which are fainter than $i = 15.0$ with equally broad absorption lines. The catalogue consists of the 46420 objects based on SDSS DR3.

### *2.2. UKIDSS*

The UKIDSS can be considered as the near-IR counterpart of the SDSS [19]. The survey will cover 7500 deg$^2$ of the Northern Sky in (Y, J, H and K) bands down to $K \cong 18.3$, thus reaching three magnitudes deeper than 2MASS [3]. There are five sub-surveys: Ultra Deep Survey (UDS), Galactic plane Survey (GPS), Galactic Cluster Survey (GCS), and Deep Extragalactic Survey (DXS) but the one of the relevance to this work is the UKIDSS Large Area Survey (LAS), whose goal is to cover 4000 deg$^2$ until 2012.

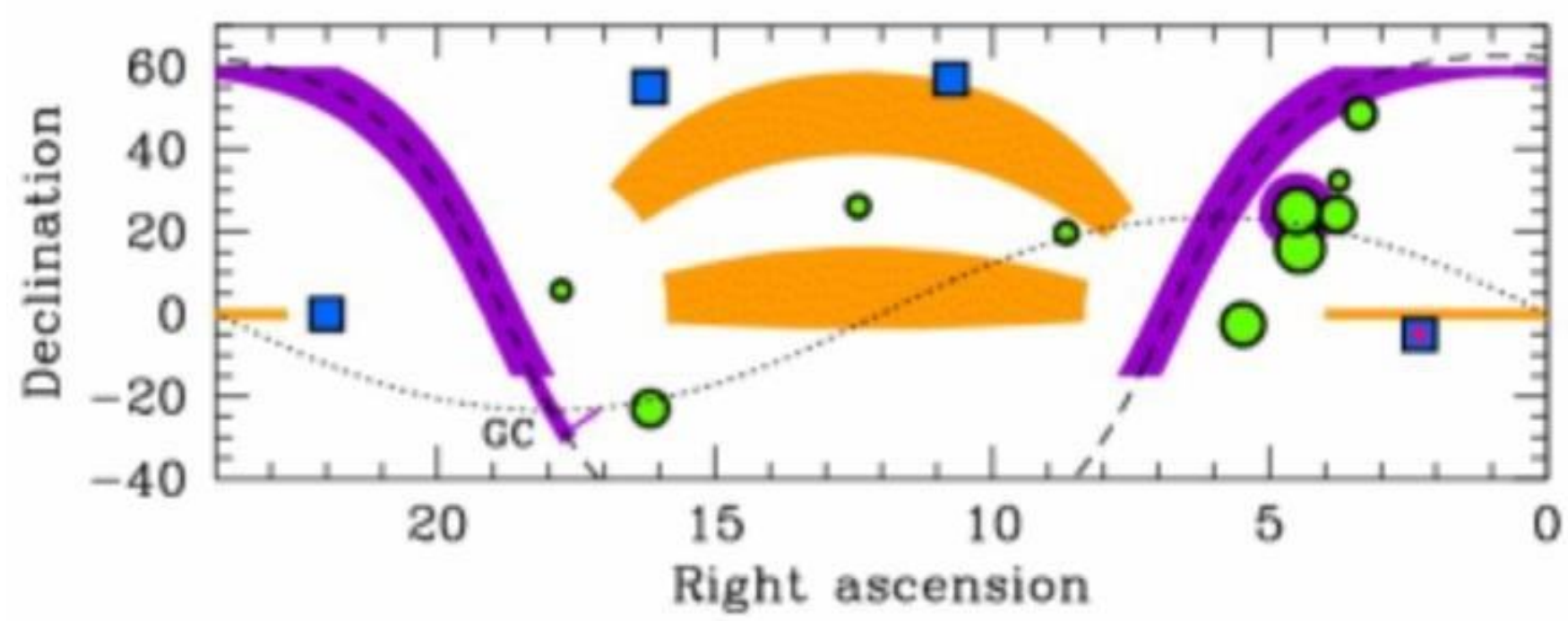

**Figure 4.** Location on the sky of the fields comprising the various survey components. Orange: LAS, Green: GCS, Violet: GPS, Blue rectangles: DXS, UDS lies just to the west of DXS field at $02^h$ $18^m$ -05° 10’. The dashed line marks the galactic plane and the dotted line marks the ecliptic. UKIRT lies at latitude +20° (Source: [33]).

The survey uses the Wide Field Camera (WFCAM) [6] which is mounted on 3.8m UKIRT. The telescope’s wide effective area of 0.21 $deg^2$ provides a relatively quick progress under tolerable seeing conditions (<1.’’2). Comprehensive descriptions of the survey hardware, software strategy and data product characteristics can be found in these papers: [19], [32].

The calibration of the zero-point for each detector in each stack multiframe is determined by identifying suitable 2MASS stars in the frame, and converting the 2MASS magnitudes to the WFCAM *JHK* system using linear colour equations [32], [8]:

$$J_{WFCAM} = J_{2MASS} - 0.075(J_{2MASS} - H_{2MASS})$$
$$H_{WFCAM} = H_{2MASS} + 0.04(J_{2MASS} - H_{2MASS}) - 0.04$$
$$K_{WFCAM} = K_{2MASS} - 0.015(J_{2MASS} - K_{2MASS})$$

The magnitudes in UKIDSS are represented in Vega system with 5 $\sigma$ accuracy and typical limiting magnitude of the LAS is JHK = [19.56, 18.81, 18.19 (Vega, 5$\sigma$)] or [20.49, 20.19, 20.09 (AB, 5$\sigma$)] (Figure 5). Uncertainty in zero-points accuracy remains present at 2 per cent level. Table 1 provides a conversion to AB magnitude system for different filters [16].

**Table 1.** Conversion to AB magnitude

| Band | $\lambda_{\text{eff}}$ (μm) | $S$ (Jy) | Flux density (W m$^{-2}$ μm$^{-1}$) | AB offset |
|---|---|---|---|---|
| $u$ | 0.3546 | 1545 | $3.66 \times 10^{-8}$ | 0.927 |
| $g$ | 0.4670 | 3991 | $5.41 \times 10^{-8}$ | −0.103 |
| $r$ | 0.6156 | 3174 | $2.50 \times 10^{-8}$ | 0.146 |
| $i$ | 0.7471 | 2593 | $1.39 \times 10^{-8}$ | 0.366 |
| $z$ | 0.8918 | 2222 | $8.32 \times 10^{-9}$ | 0.533 |
| $Z$ | 0.8817 | 2232 | $8.59 \times 10^{-9}$ | 0.528 |
| $Y$ | 1.0305 | 2026 | $5.71 \times 10^{-9}$ | 0.634 |
| $J$ | 1.2483 | 1530 | $2.94 \times 10^{-9}$ | 0.938 |
| $H$ | 1.6313 | 1019 | $1.14 \times 10^{-9}$ | 1.379 |
| $K$ | 2.2010 | 631 | $3.89 \times 10^{-10}$ | 1.900 |

In the Table 1 $\lambda_{eff}$ is effective wavelength, columns 3 and 4 list the flux density of an object spectrum and 5th column shows the AB magnitude offset when converting the Vega-based magnitudes to AB magnitude system using following definition:

$$AB_\nu = -2.5 log f_\nu + 48.6$$

Astronomical accuracy in the LAS is typically < 0.''1 and photometric accuracy is ~ 0.04 magnitudes (in J-band).

The processed data is accumulated in WFCAM Science Achieve (WSA) [14] through automated pipeline. Hewett et al. (2006) explains that before the frames are catalogued into the database, the pipeline flat-fields the data, subtracts the counts from the background sky, detects and parameterises objects and performs the photometric and astrometric calibrations. Finally, the data can be accessed through a flexible query tool, which allow Structured Query Language (SQL) commands for sophisticated searches.

Here is a short example of a typical SQL for WSA (SDSS supports similar structure):

```
select top 100
ra,dec,l,b,jAperMag3,jAperMag3Err,hAperMag1,hAperMag1Err,k_1AperMag3,k_1AperM
ag3Err,pStar
from UKIDSSDR6PLUS..gpsSource where ra > 1 and ra < 120 and dec > -10 and dec
< 10
and jAperMag3 > 0 and hAperMag3 > 0 and k_1AperMag3 > 0
and pStar > 0.99
and mergedClass !=0 and (PriOrSec=0 or PriOrSec=framesetID)
```

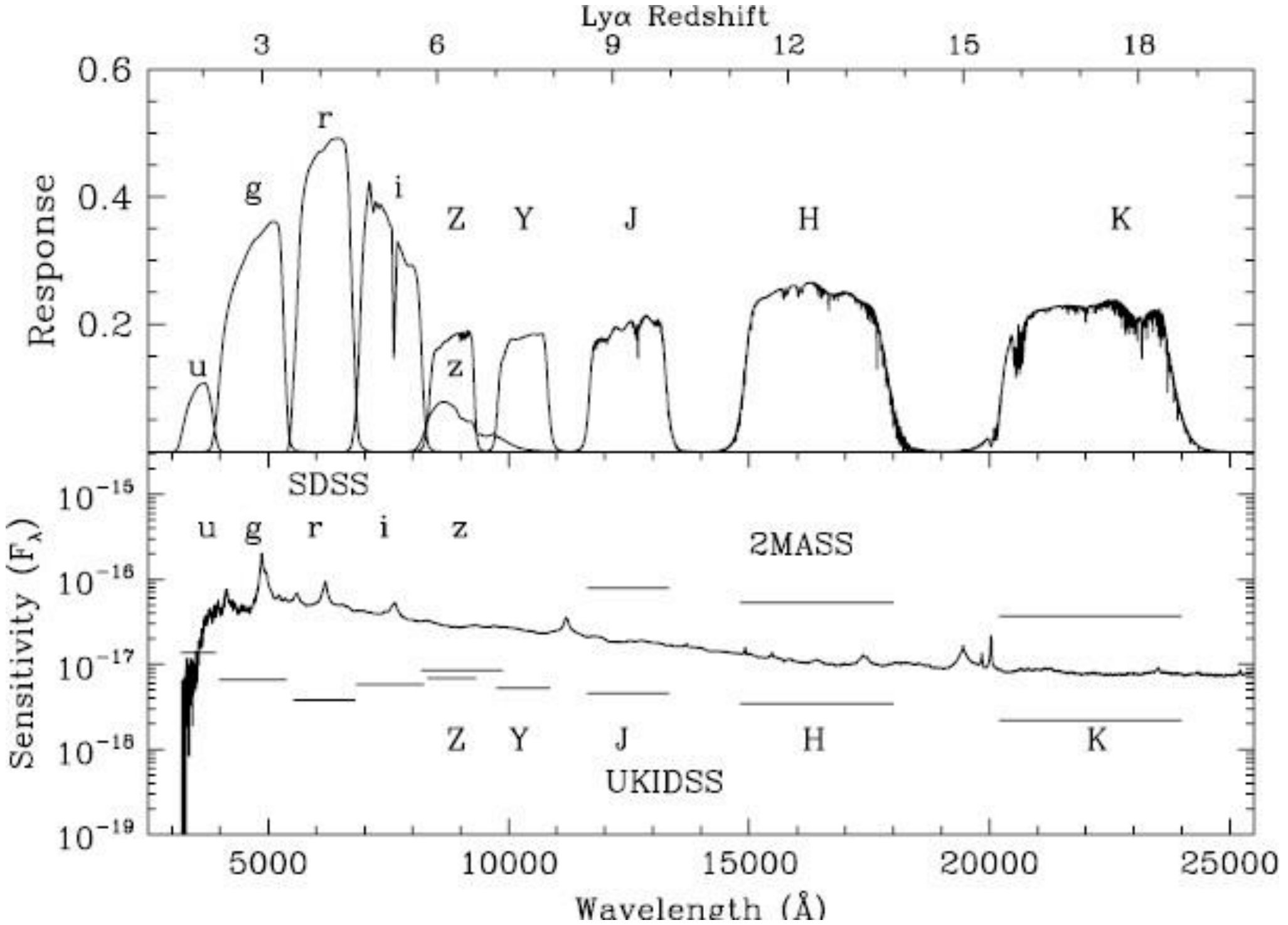

**Figure 5.** Top panel: throughputs of the SDSS and UKIDSS imaging systems. The new UKIDSS Z an Y passbands have not been used in this project, they have sharper cut-offs compared to SDSS *z*, and fill the wavelength gap between the SDSS and NIR JHK bands. For reference, the top axis scale indicates the wavelength position of quasar Ly*α* emission versus redshift. Bottom panel: the horizontal lines indicate the resulting $5\sigma$ $F_\lambda$ sensitivities (erg $s^{-1}cm^{-2}Å^{-1}$) and wavelength coverage of the SDSS and UKIDSS survey bands, in addition to the shallower 2MASS in JHK. A composite quasar spectrum shifted to redshift z=3 and normalised to magnitude $i_{AB}$ = 19.1 is shown, demonstrating NIR detection in the UKIDSS (Source: [7]).

# 3. QUASAR COLOURS

## *3.1. Matching Procedures*

SDSS – UKIDSS quasar positions overlap in the northern and southern areas of the sky as well as in a significant portion of the southern equatorial stripe. The common area extends from a right ascension range of approximately $22^h$ to $4^h$ and a declination range of ±1.265°, centred on 0°. Extracted quasars from the SDSS DR7 catalogue with a redshift constraint of $0 < z < 3$ and classification ‘QSO’ then were positionally cross-matched with he UKIDSS DR6 database using tolerance of 1’’.0 and selecting only the nearest object to the queried position. A total of 5730 objects have been successfully matched in the SDSS – UKIDSS system (Figure 6).

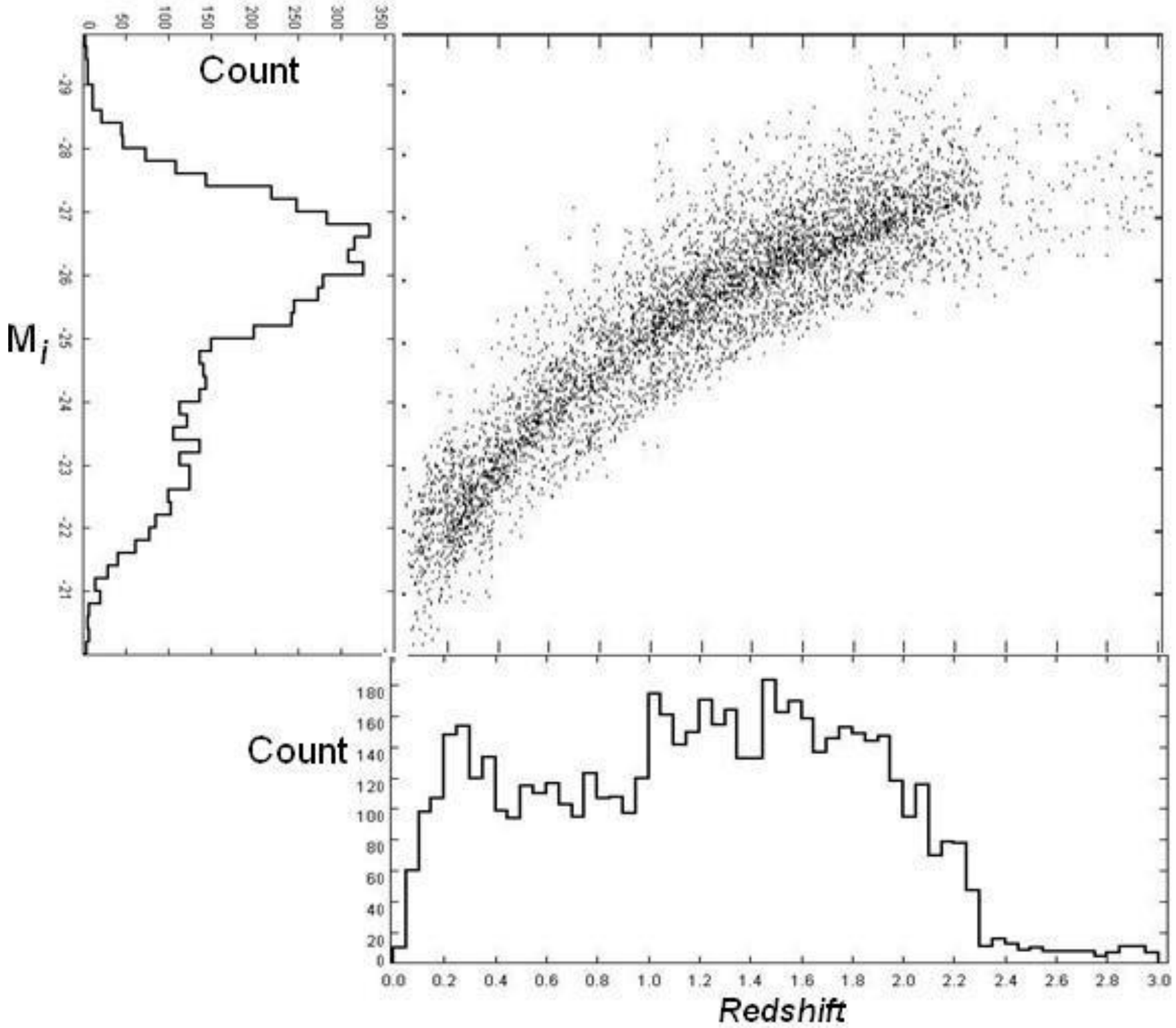

**Figure 6.** The redshift and absolute $M_i$ magnitude distribution of 5730 SDSS-UKIDSS matched quasars is shown in the central panel. There is a density edge is noticeable due to the magnitude limits of the SDSS main quasar spectroscopic survey. The side panels illustrate the distributions of absolute magnitude (AB system) and redshift. Only intrinsically bright quasars can be seen at far redshifts.

In Figure 6, there is a sharp cut-off of the quasar distribution over redshift at z ~ 2.3. The observed quasar population beyond z ~ 2.3 is affected by the declining space density of the objects and small volumes probed. As a result, the flux-limited sample becomes quite sparse. At high redshift, the luminosity magnitude of a sample causes a selection effect such only objects that have intrinsically high luminosity are found. However, at low redshifts, intrinsically faint quasars are just bright enough to make it into the sample. Thus, the selection criterion is an effect based on absolute magnitude which marks the prominent edge on the redshift [7].

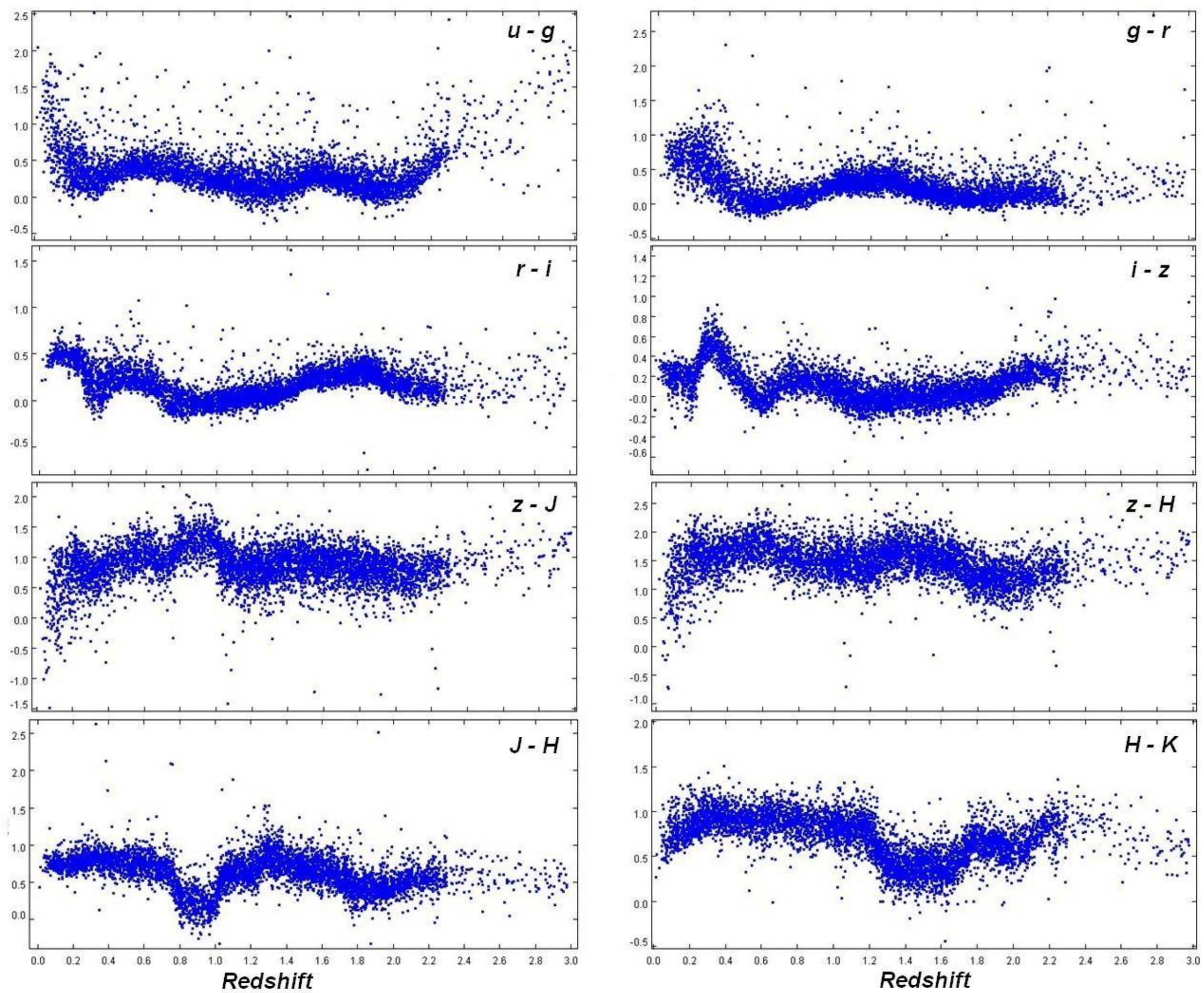

**Figure 7.** Colours of 5730 quasars from SDSS-UKIDSS system illustrated in *ugriz*JHK combinations of interest versus redshift. The plots progress from optical to NIR colours and their magnitudes are calculated in AB magnitude system.

### *3.2. Quasar Colours versus Redshift*

In Figure 7, 5730 SDSS-UKIDSS quasars are plotted in various colours versus SDSS spectroscopic redshift. Each colour-plot only contains quasars with detected flux in both contributing bands, but objects with non-detections in the necessary filters with magnitudes indicated -9999.99 in either SDSS or UKIDSS are omitted. The colour behaviour with redshift is thoroughly analysed in Richards et al. (2002), Maddox & Hewett (2006) and Richards et al. (2006a). Chiu et al. (2007) noted that the passage of the strong emission lines in and out of various filters controls the small-scale fluctuations of the quasar colours versus redshift, while the extended underlying continuum provides the distinguishing characteristics separating quasar colours from those of stars. The strongest observed broad emission lines are the hydrogen Balmer-series lines (H$\alpha$ $\lambda$6563, H$\beta$ $\lambda$4861, H$\gamma$ $\lambda$4340), hydrogen Ly$\alpha$ $\lambda$1216 and prominent lines of abundant ions (MgII $\lambda$2798, CIII $\lambda$1909 and CIV $\lambda$1549). According to Peterson (1997), these lines appear in most quasar spectra, but depending on the redshift of the quasar, some may not be observable if they fall outside the spectral widow of a particular filter. Typical flux ratios are given in Table 2, along with equivalent widths.

**Table 2. E**mission line strengths in AGN [25]

| Line | Relative Flux (Ly$\alpha$ + NV = 100) | Equivalent Width (Å) |
|---|---|---|
| Ly$\alpha$ $\lambda$1216 + NV $\lambda$1240 | 100 | 75 |
| CIV $\lambda$1549 | 40 | 35 |
| CIII $\lambda$1909 | 20 | 20 |
| MgII $\lambda$2798 | 20 | 30 |
| H$\gamma$ $\lambda$4340 | 4 | 30 |
| H$\beta$ $\lambda$4861 | 8 | 60 |

Fan X (1999) predicted the evolution of quasar colours with redshift using Monte Carlo simulation in SDSS bands. Those simulations show that for $z < 2.0$, the $u - g$ colour of quasars remain near 0.3. In that redshift range, quasar colours are dominated by the power-law continuum. The colours of a pure power-law spectrum are independent of redshift for a given index $\alpha$. It predicts that all the emission lines, except Ly$\alpha$, have insufficient strength to affect the broadband colours. At $z \sim 2.0$, Ly$\alpha$ coincides with u band and pushes $u - g$ to blue side. When Ly$\alpha$ emission moves to g filter at $z \sim 2.5$, the $u - g$ colour becomes redder. Lyman absorption systems absorb most of the radiation in u band at $z > 2.5$ and $u - g$ reddens rapidly. The evolution versus redshift of the redder colours (g-r, r-i, i-z) follows the same route but shifted to the higher redshifts. The model does not predict the behaviour of the broad absorption lines (BALs), however, the simulated colours agree with observations fairly well [9].

One of the first attempts to analyse quasars in IR regime was found in the work by Hatziminaoglou et al. (2005) where 35 optically confirmed quasars were explored by Spitzer Wide-Area Infrared Extragalactic Survey (SWIRE). The observed plots nearly match the ones produced in this work, but clearly does not show the entire picture due to the lack of statistical data. Chiu et al. (2007) illustrates (Figure 8) the composite quasar spectrum constructed from UV to NIR for high-luminosity quasars over redshifts $0.5 < z < 2$. Care was taken to exclude the host galaxy contamination on the observed properties of near quasars [22].

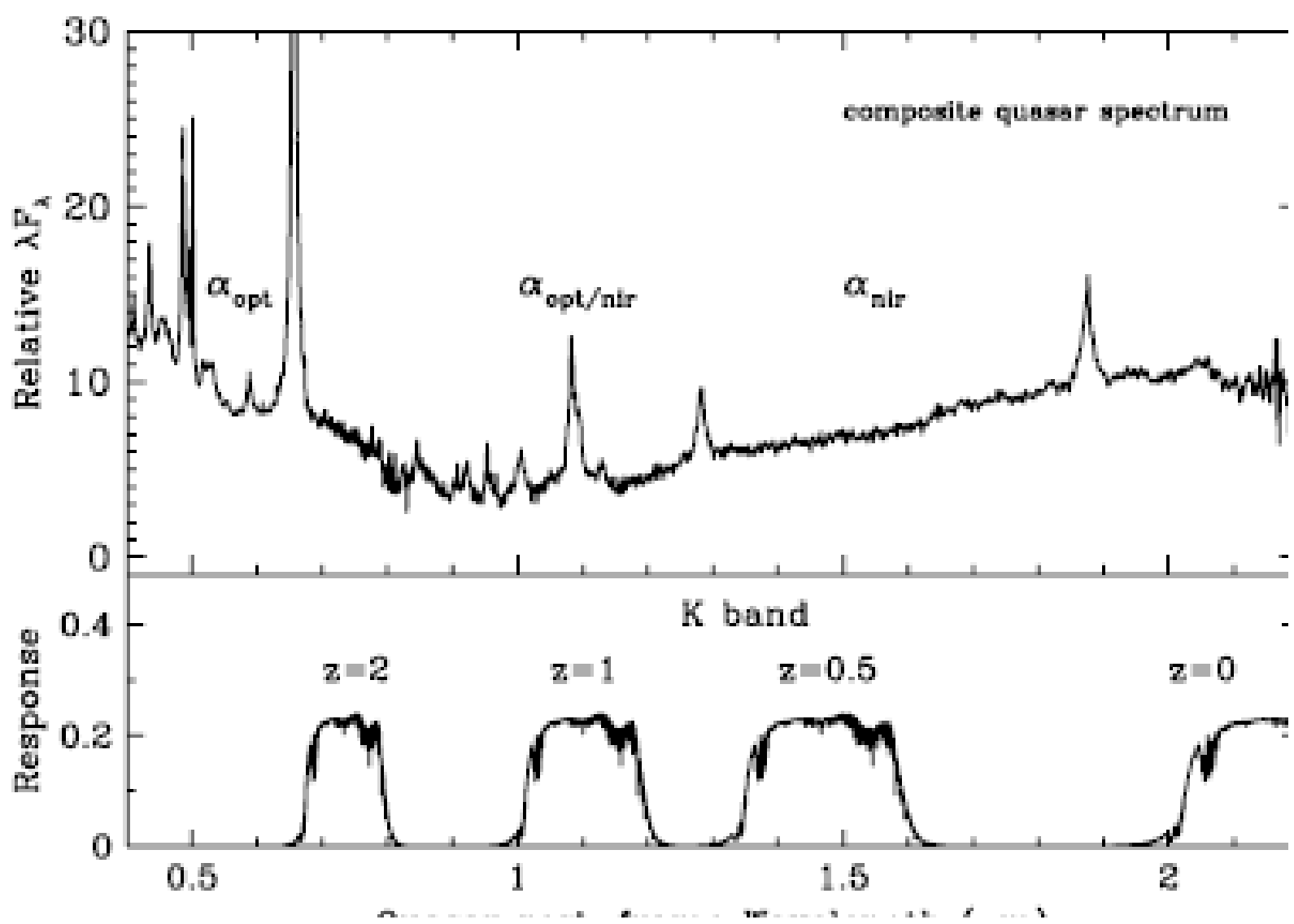

**Figure 8.** As quasars of increasing redshift are selected using K band excess (KX) – technique, the K filter probes increasingly blue rest-frame wavelengths of the quasar SEDs. As a result, different continuum slope regions ($\alpha_{opt/NIR}$) are sampled and it is important that appropriate K-corrections are incorporated in any analysis of the properties of the target populations. A rest-frame quasar composite spectrum derived from Chiu et al. (2007).

### *3.3. K-corrections*

K-corrections is defined [11] as the difference between the rest-frame absolute magnitude, M, of a source and its observed magnitude, *m*, at redshift *z*, through the same band-pass. Mathematically it can be expressed as:

$$m = M + 5\log_{10}\left(\frac{D_L}{10pc}\right) + K(z)$$

where $D_L$ is the luminosity distance. The K-correction is then computed from the source flux per unit wavelength, $f(\lambda)$ and the response function for the band-pass, $T(\lambda)$ with the following equation:

$$K(z) = 2.5\log_{10}(1+z) + 2.5\log_{10}\left[\frac{\int T(\lambda)f(\lambda)d\lambda}{\int T(\lambda)f\left(\frac{\lambda}{1+z}\right)d\lambda}\right]$$

### *3.4. Colour-colour diagrams*

**Figure 9.** SDSS-UKIDSS quasars (small dots) in colour-colour diagram (*ugriz*JHK) with indication of the redshift. Most objects are located in small regions but some do spread over wide ranges, especially in u-g colour. Colours here are calculated with a conversion to the AB magnitude system.

Figure 9 illustrates the colour versus colour plots of the quasars of the SDSS-UKIDSS system. These observational diagrams mainly confirm the simulated model which is described in Richards et al. (2006a). The simulations reproduce the overall trend in the relative colours, but some deviations are present, in particularly prominent in u-g colour. This is good indication of a departure from the power law. Quasars are point sources and therefore it is easy to confuse them with normal stars. According to Richards et al. (2006b), colour-colour plots are very powerful tools to distinguish quasars from other objects. Particularly, g-r versus u-g as well as H-K versus J-H diagrams are used in selection techniques. Considering the future interest, combination of NIR with mid-IR imaging from the *Spitzer Space Telescope* [15] may potentially reach new distant and faint quasar populations while providing nearly continuous band-pass coverage as emission lines and continuum flux shifts with redshift.

# 4. DISCUSSIONS

## *4.1. Quasar Reddening*

Scatter increases towards the red with fainter magnitude on the first plot of Figure 9 (blue colours). It is likely to be the result of reddening of quasars, either external or internal, but certainly extragalactic. The fact that the reddening is independent of redshift points toward internal reddening, since external one would depend on the volume of space, hence the redshift.

If unified models are correct and accretion processes form non-spherical structures to fuel the quasars, then the presence of an obscured population of quasars is a natural consequence of the present understanding of AGN models, because the line-of-sight to the inner regions of AGN will be certainly blocked by gas and dust. Highly reddened quasars with u-g > 2 have been estimated to be more than 30 objects which is small fraction of the population. Indeed, examples of obscured objects have been identified, whose spectra show small deviations consistent with obscuration by dust at UV wavelengths [27]. Since these objects are proven to exist, the question of what fraction of quasar population is reddened naturally arises. Also there is little know about intergalactic reddening and more research is needed to understand if internal reddening is common phenomenon.

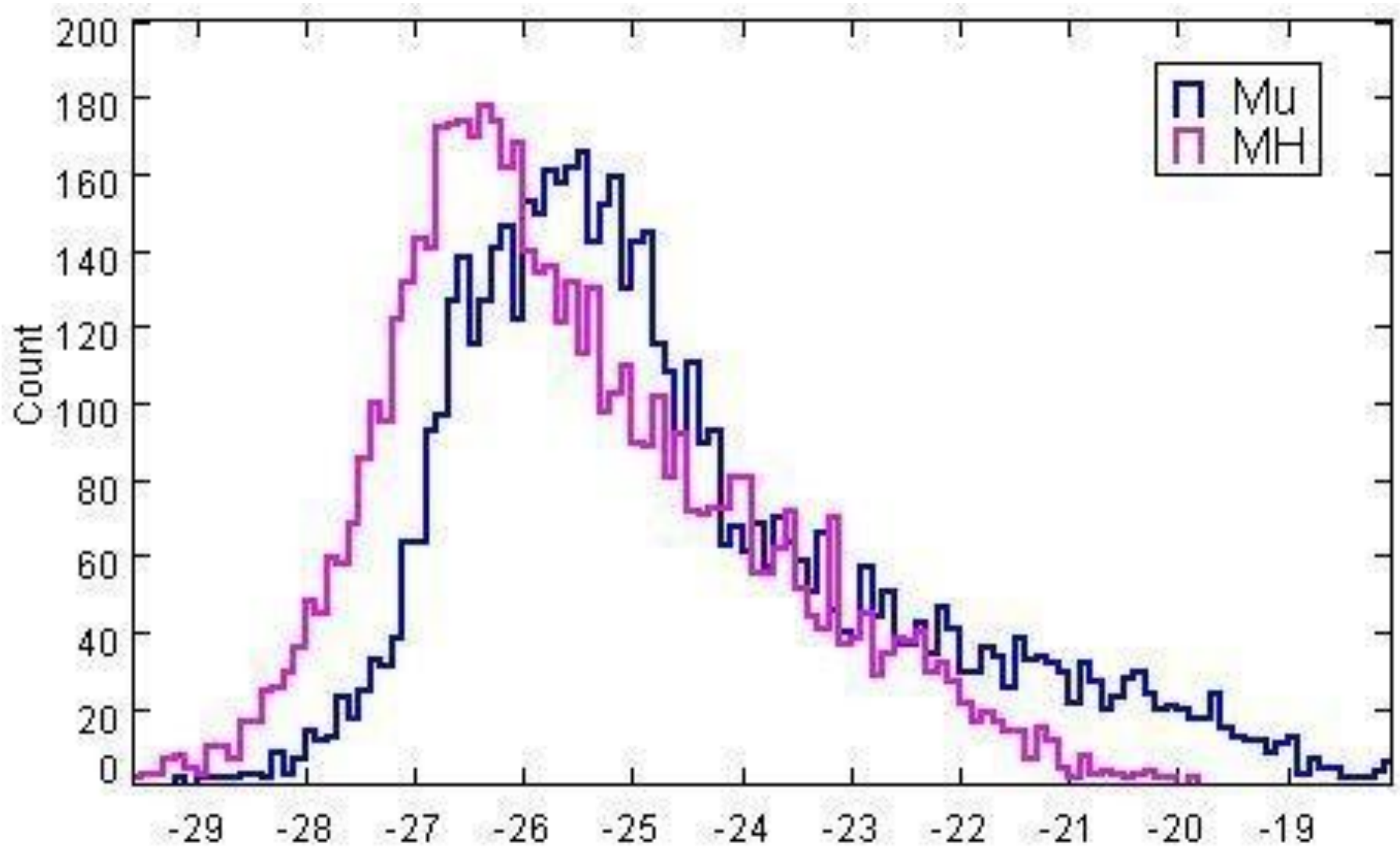

**Figure 10.** The histogram is absolute magnitude distribution of the SDSS-UKIDSS quasars in *u* and H bands. Magnitude bins have a width of 0.1. If the fall in brightness of the *H* colour is steady, *u* band produces a flat feature in lower brightness. It is an indication of the reddening of quasars.

### *4.2. Notes on individual objects*

**SDSS J134031.78+000022.7** (z = 0.012): This object has very low activity in the nucleus. No broad line emissions are seen from the spectra. It is located nearby and extended shape is observable. It is more likely to be a normal galaxy rather than an AGN.

**SDSS J153702.77+062203.8** (z = 2.503): A BAL with a number of quite impressive absorption troughs

**SDSS J154751.93+025550.8** (z = 0.098): The highest possible u-g colour > 6.0, well resolved emission lines OIII, SII and NII.

**SDSS J093411.51+002952.0** (z = 1.918): Shows very high extinction in NIR (J-H) colour.

**SDSS J091442.32+000637.1** (z = 0.561): Highly reddened in u band (2 magnitudes). Emission spectra does not have broad emission lines, instead OIII lines are resolved clearly. The flux does not seem to obey power law, but is kept constant in optical.

**SDSS J115448.72+094837.8** (z = 0.019): This object possibly is not a QSO. It has low magnitudes in all bands and there are no broad lines in spectrum. It is more appear to be a normal galaxy.

### *4.3. Other Selection Methods for Quasars*

In this work quasars were readily identified by SDSS algorithms. Some peculiar objects at extreme edges on the colour diagrams were questioned on its classification (Figure 9). These findings suggest that the SDSS system's quasar selection methods should be checked for consistency. Furthermore, a common distinction at ($M_V$ = - 23.0) between Seyfert galaxies and QSO would be useful to include to the SDSS. However, it is not the goal of the current paper to discuss the quasar selection algorithm of the SDSS in great detail. Colour-colour diagrams with K pass-band showed that NIR bands can be similarly used to find quasars and enrich the catalogues. Maddox et al. (2008) discusses the quasar population which is selected in KX sample that do not appear in the SDSS catalogue.

## 5. CONCLUSION

The result of this work is a population of 5730 quasars selected from the SDSS DR7 and UKIDSS DR6. An empirical investigation of quasars ($0 < z < 3$) from UV/optical to near-IR bands were carried out to test the current models of AGN. The distribution of quasars was critically analysed based on (1) a quasar spectral model, which includes the observed distributions of power-law continuum, optical/UV emission lines and a large number of absorption systems due to intervening neutral hydrogen; and (2) the evolution of the colour functions with redshift. The colours of quasars with $2.5 < z < 3.0$ are possible to confuse with hot stars. The results also indicated that there was a point of highest activity of quasar generation at $z \sim 2.0$. However, the surveys have their magnitude limits which do not allow observing fainter quasars. From u-g colour plots it was found that quasars undergo reddening due to internal obscuration by dust and gas. This point has been checked and confirmed by analyzing the absolute magnitude in blue colours, where the decrease in intrinsic brightness yielded a flat feature.

# APPENDIX

These spectra of quasars are taken from SDSS DR7.

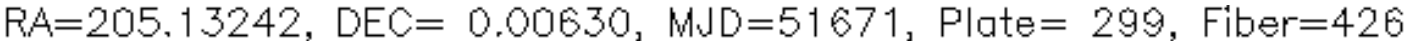

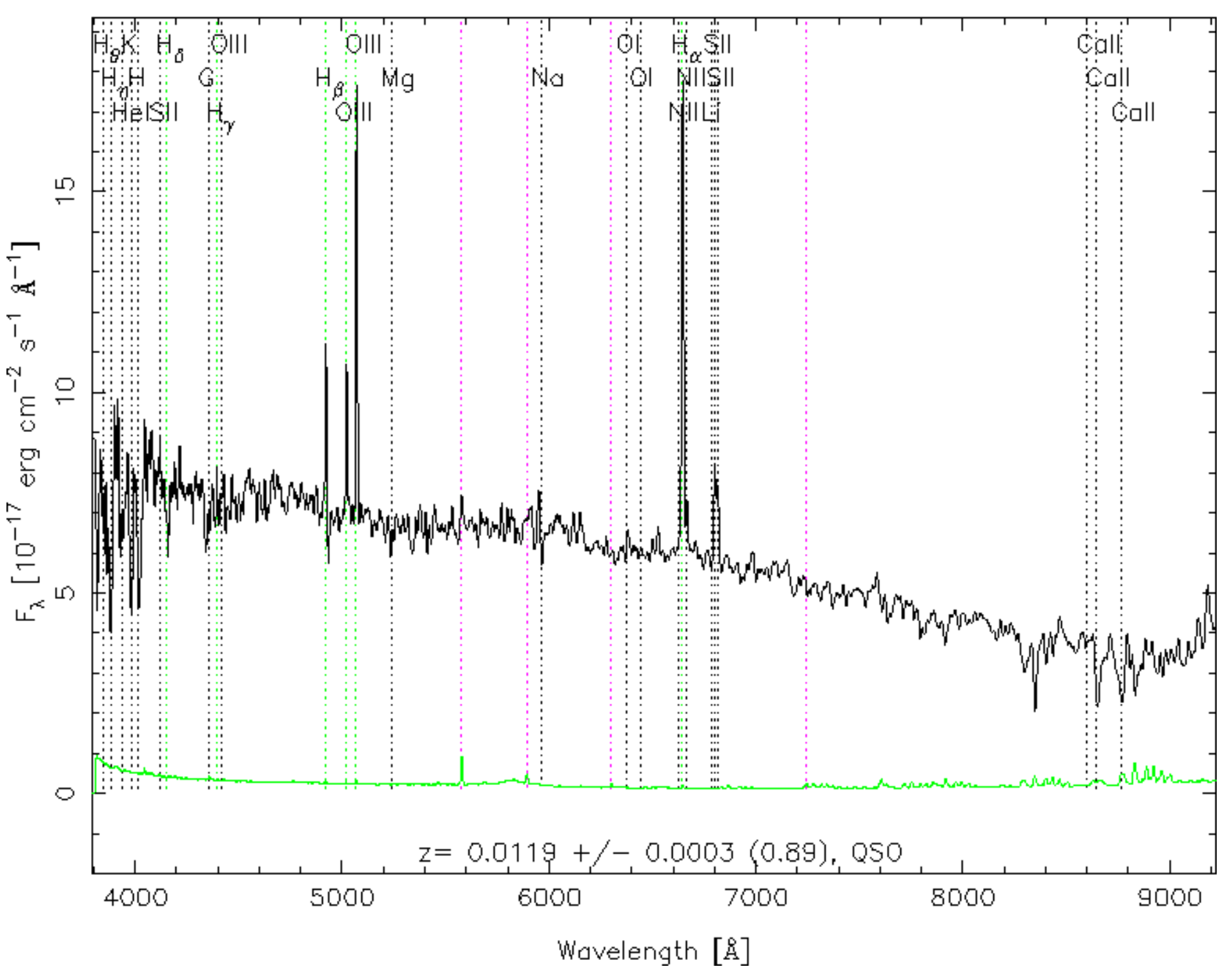

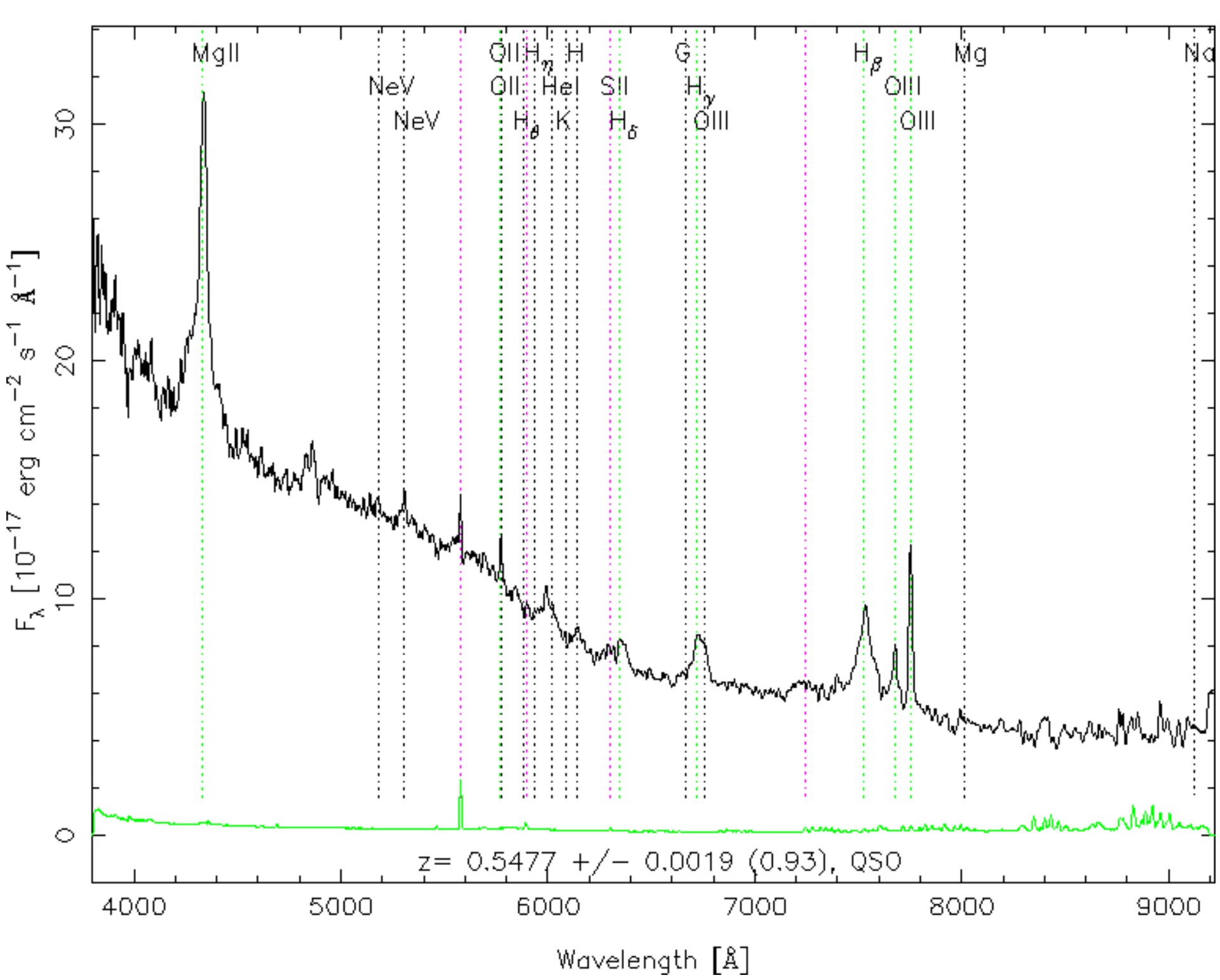

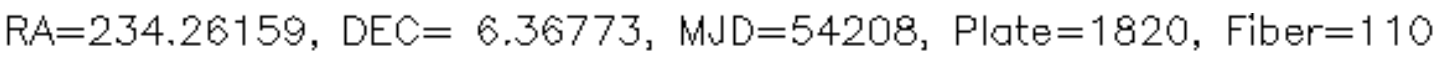

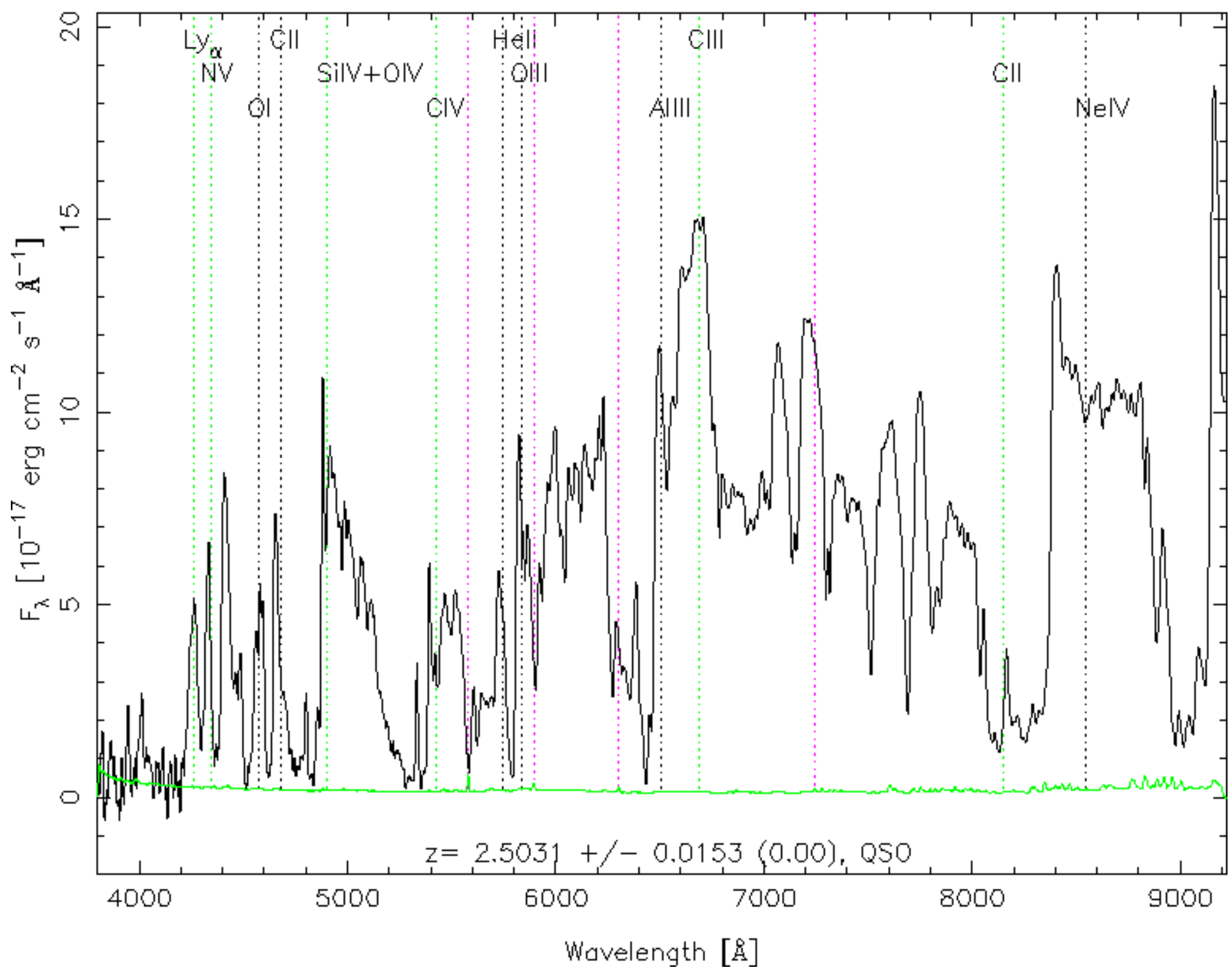

RA=236.96642, DEC= 2.93079, MJD=52045, Plate= 594, Fiber= 86

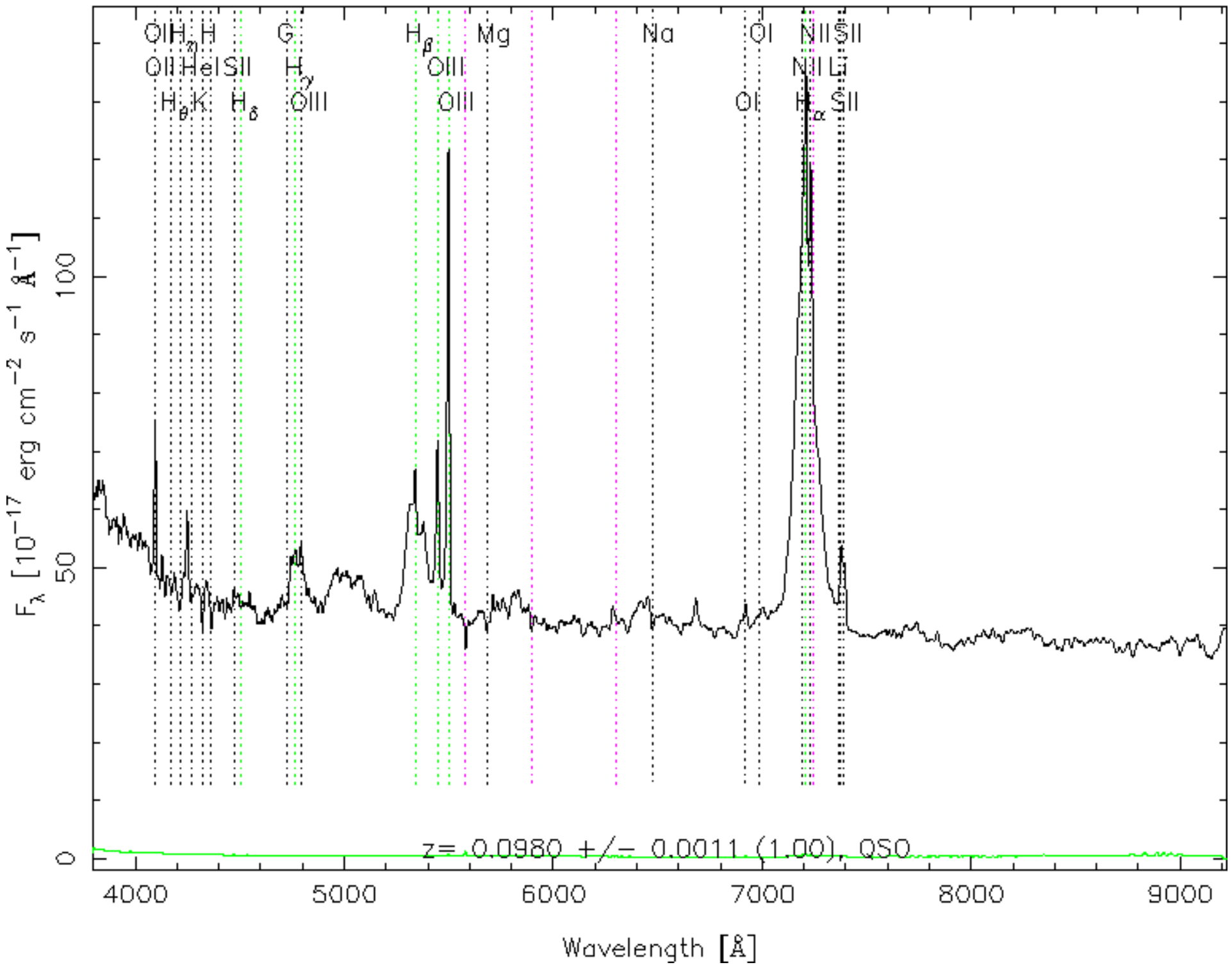

RA=138.67633, DEC= 0.11032, MJD=51955, Plate= 472, Fiber=161

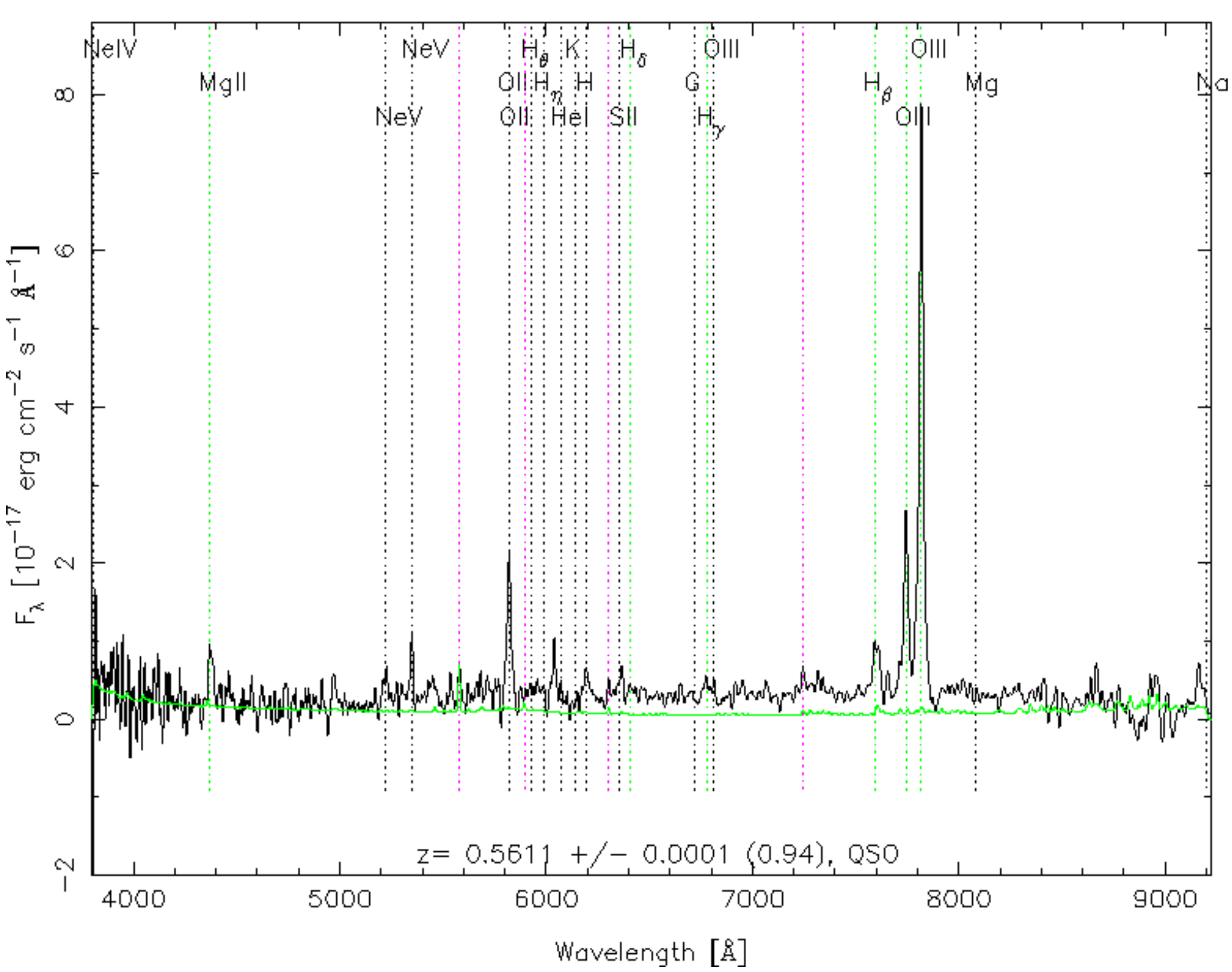

RA=178.70302, DEC= 9.81049, MJD=52733, Plate=1227, Fiber=176

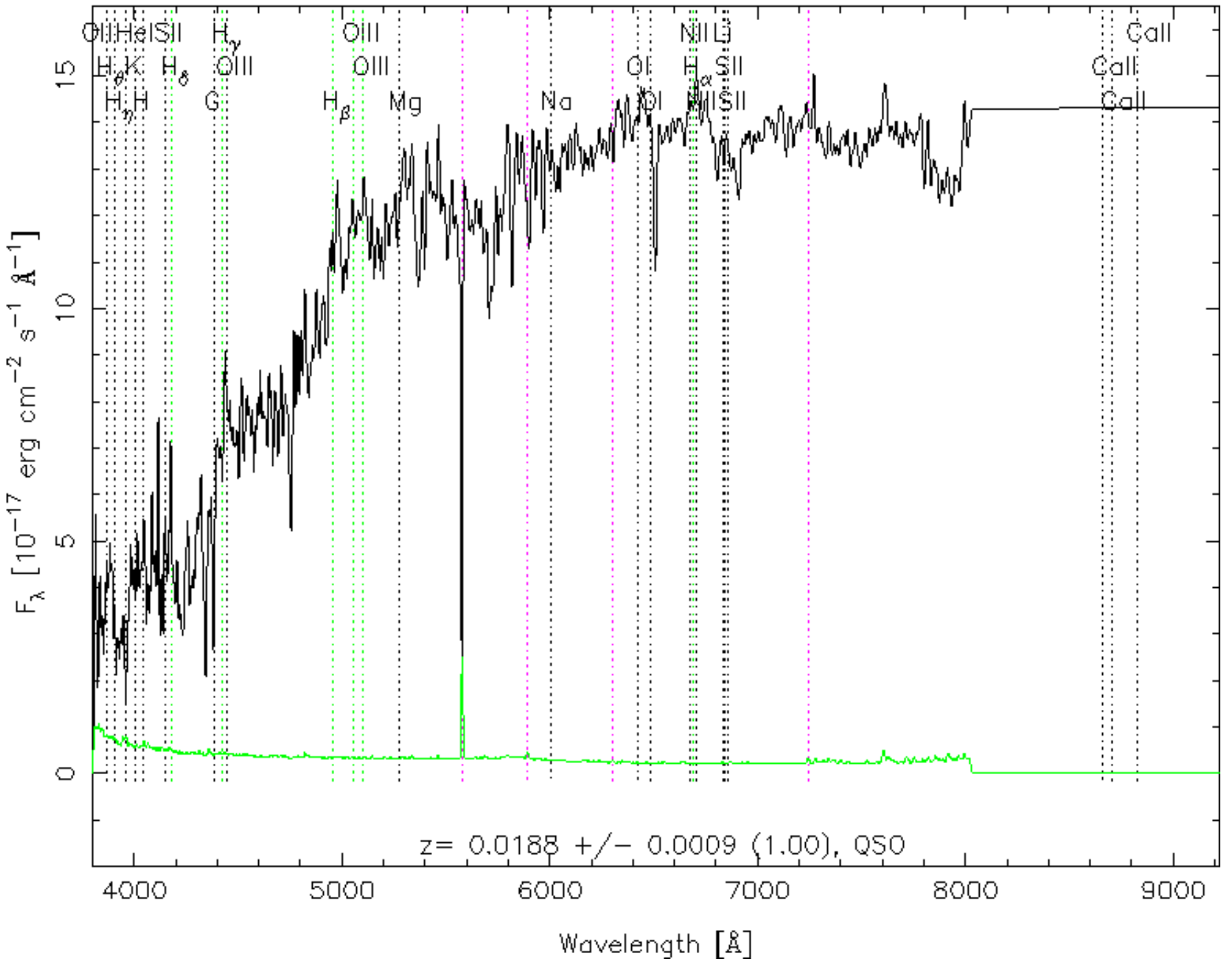